\documentclass[12pt]{iopart}
\usepackage{iopams}

\usepackage{graphicx}

\usepackage[normalem]{ulem}
\usepackage[dvipsnames]{xcolor}
\usepackage{upgreek}
\usepackage{hyperref}

\usepackage{cite}

\usepackage{xcolor}

\begin{document}

\title{Molecule-molecule and atom-molecule collisions with ultracold RbCs molecules}

\author{$^{1}$Philip D. Gregory, $^{1}$\footnote{Present address: Department of Physics, University of Oxford, Oxford, United Kingdom, OX1 3PU.}Jacob A. Blackmore, $^{2}$Matthew D. Frye, $^{1}$Luke M. Fernley, $^{1}$Sarah L. Bromley, $^{2}$Jeremy M. Hutson, and $^{1}$Simon L. Cornish}

\address{$^{1}$Joint Quantum Centre (JQC) Durham-Newcastle, Department of Physics, \\ Durham University, Durham, United Kingdom, DH1 3LE.}
\address{$^{2}$Joint Quantum Centre (JQC) Durham-Newcastle, Department of Chemistry, \\ Durham University, Durham, United Kingdom, DH1 3LE.}
\vspace{10pt}
\date{\today}
\begin{abstract}

Understanding ultracold collisions involving molecules is of fundamental importance for current experiments, where inelastic collisions typically limit the lifetime of molecular ensembles in optical traps. Here we present a broad study of optically trapped ultracold RbCs molecules in collisions with one another, in reactive collisions with Rb atoms, and in nonreactive collisions with Cs atoms. For experiments with RbCs alone, we show that by modulating the intensity of the optical trap, such that the molecules spend 75\% of each modulation cycle in the dark, we partially suppress collisional loss of the molecules. This is evidence for optical excitation of molecule pairs mediated via sticky collisions. We find that the suppression is less effective for molecules not prepared in the spin-stretched hyperfine ground state. This may be due either to longer lifetimes for complexes or to laser-free decay pathways. For atom-molecule mixtures, RbCs+Rb and RbCs+Cs, we demonstrate that the rate of collisional loss of molecules scales linearly with the density of atoms. This indicates that, in both cases, the loss of molecules is rate-limited by two-body atom-molecule processes. For both mixtures, we measure loss rates that are below the thermally averaged universal limit.  

\end{abstract}

Ultracold polar molecules have been proposed for applications in the fields of quantum computing~\cite{DeMille2002, Yelin2006, Zhu2013, Herrera2014, Ni2018, Sawant2020, Hughes2020}, quantum simulation~\cite{Barnett2006, Micheli2006, Buchler2007, Gorshkov2011, Macia2012, Manmana2013, Gorshkov2013}, quantum-state-controlled chemistry~\cite{Heazlewood2021,Krems2005,Krems2008,Balakrishnan2016}, and precision measurements~\cite{DeMille2017,Safronova2018,Tarbutt2013,Aggarwal2018,Hutzler2020,Augenbraun2020,Zelevinsky2008,Schiller2014}. Experiments are now able to produce a wide variety of ultracold polar molecules in the ground state by association of atom pairs in a pre-cooled atomic mixture~\cite{Ni2008, Takekoshi2014, Molony2014, Park2015, Guo2016, Rvachov2017, Seesselberg2018, Yang2019, Voges2020, Cairncross2021} or by direct laser cooling of the molecules~\cite{Shuman2010, Hummon2013, Zhelyazkova2014, Barry2014, McCarron2015, Norrgard2016, Kozyryev2017, Truppe2017, Lim2018, Anderegg2018, Ding2020}. 
In these experiments, the densities of the gases are often high enough that the effects of collisions are measurable and typically limit the trap lifetime of the molecular gas.
Understanding ultracold collisions involving molecules is both fundamentally interesting, due to their complexity, and crucial for further developing the techniques needed to control collisional losses during experiments~\cite{Zhu2013zeno, Karman2018, Lassabliere2018, Karman2019, Matsuda2020, Li2021, Anderegg2021}.

All ultracold molecules investigated to date experience fast collisional losses from optical traps characterised by second-order kinetics~\cite{Ospelkaus2010, Ni2010, Rvachov2017, Cheuk2020, Takekoshi2014, Park2015, Ye2018, Guo2018, Gregory2019, Gersema2021, He2021, Bause2021}. The molecules can be broadly categorised as either reactive or nonreactive in photon-free two-body collisions; 
for a pair of identical nonreactive molecules ($XY$) composed of the atomic species $X$ and $Y$, the atom-exchange reactions of the form
\begin{equation*}
2XY\rightarrow X_2 + Y_2,
\end{equation*}
\begin{equation*}
2XY\rightarrow X_2 Y + Y,
\end{equation*}
\begin{equation*}
2XY\rightarrow X + Y_2 X 
\end{equation*}
are all endothermic~\cite{Zuchowski2010}. In contrast, for reactive molecules at least one of these reactions is exothermic. Accordingly, these atom-exchange reactions can cause fast loss of reactive molecules, but not nonreactive molecules. 
Despite this, two-body collision rates at or near the universal limit~\cite{Idziaszek2010, Frye2015}, where all molecules that reach the short-range part of the interaction potential are lost, have been observed in all experiments so far, independent of whether the molecules are reactive~\cite{Ospelkaus2010, Ni2010, Rvachov2017, Cheuk2020} or nonreactive~\cite{Takekoshi2014, Park2015, Ye2018, Guo2018, Gregory2019,Yang2019, He2021,Gersema2021,Bause2021}. 
In particular, Ye~\emph{et al.}~\cite{Ye2018} have compared the loss of $^{23}$Na$^{87}$Rb molecules in ground and first-excited vibrational states and confirmed high loss and heating rates regardless of the energetics of the exchange reactions.

A two-step process has been proposed to explain the fast loss of nonreactive molecules from optical traps. First, during the collision a long-lived two-molecule collision complex is formed~\cite{Mayle2012, Mayle2013, Christianen2019DOS, Christianen2019}; this is commonly referred to as a `sticky collision'. The lifetime of the complex is commonly estimated as
\begin{equation}
\tau_\mathrm{c}=\frac{2\pi\hbar\rho}{N_0},
\end{equation}
where $\rho$ is the density of the accessible rovibrational states and $N_0$ is the number of open channels, with $N_0=1$ for nonreactive molecules in their absolute ground states.  
This is based on Rice-Ramsperger-Kassel-Marcus (RRKM) theory~\cite{Levine2005}, which effectively assumes that the motion is ergodic, \textit{i.e.}\ that energy is fully randomised in the complex. Second, once the complex is formed it may be removed from the trap due to electronic excitation by the trap light leading to permanent loss of the molecule pair from the sample~\cite{Christianen2019DOS}.

We have previously established the validity of this two-step loss process for RbCs molecules. We demonstrated that loss of RbCs molecules from an optical trap is rate-limited by a two-body process~\cite{Gregory2019}, consistent with the `sticky collision' hypothesis that the loss is mediated by the formation of long-lived two-molecule collision complexes. More recently, we showed that optical excitation is the dominant mechanism by which the (RbCs)$_2$ complexes are removed from the gas~\cite{Gregory2020}. We observed a suppression of the collisional loss by applying square-wave modulation to the intensity of the optical trap to form a time-averaged potential where 75\% of each modulation cycle is dark. By varying the frequency of the modulation, we measured a lifetime for the complex in the dark of 0.53(6)\,ms~\cite{Gregory2020}, within a factor of $\sim2$ of the RRKM prediction of 0.253\,ms~\cite{Christianen2019DOS}. A similar level of agreement between theory and experiment has been found for the lifetime of the (KRb)$_2$ complex formed in reactive KRb+KRb collisions~\cite{Liu2020}.
However, experiments in intensity-modulated traps with nonreactive $^{23}$Na$^{39}$K, $^{23}$Na$^{40}$K, and $^{23}$Na$^{87}$Rb have observed no suppression of the collisional loss~\cite{Bause2021, Gersema2021}. Moreover, fast losses also exist in experiments with $^{23}$Na$^{40}$K confined to a repulsive box potential, where the molecules spend the majority of the time in the dark~\cite{Bause2021}. These results indicate either much faster optical excitation rates than expected or sticking times that are at least an order of magnitude greater than the RRKM prediction.

Atom-molecule collisions offer a compromise between the relative simplicity of collisions between alkali-metal atoms and the complexity of collisions between diatomic molecules. For molecules formed by association, it is convenient to study collisions between the associated molecule and the constituents of the initial atomic mixture. In this case, atom-molecule combinations $(X+XY)$ can be defined as nonreactive if the atom-exchange reaction
\begin{equation*}
X+XY \rightarrow X_2 + Y
\end{equation*}
is endothermic. However, as was found for molecule-molecule collisions, the reactivity of the combination alone does not appear to determine the rate of collisional loss observed in experiments~\cite{Ospelkaus2010, Yang2019, Son2020, Jurgilas2021, Jurgilas2021b, Nichols2021, Son2021, Voges2021}. For example, reactive atom-molecule collisions in mixtures of triplet $^{23}$Na$^{6}$Li molecules and $^{23}$Na atoms are sufficiently suppressed for the fully stretched hyperfine states that efficient sympathetic cooling of the molecules is possible~\cite{Son2020}. Experiments studying nonreactive $^{23}$Na$^{39}$K+$^{39}$K collisions have measured a hyperfine-dependent two-body loss rate far below the universal limit~\cite{Voges2021}. However, Nichols~\emph{et al.}~\cite{Nichols2021} recently measured a photon-free lifetime for complexes in nonreactive collisions for a  $^{40}$K$^{87}$Rb+$^{87}$Rb mixture to be $\sim10^{5}$ times higher than predicted by RRKM; it is therefore possible that optical excitation of long-lived two-body collision complexes also plays a role in the fast losses observed in other atom-molecule collisions. Yang~\emph{et al.}~\cite{Yang2019} have reported the existence of Feshbach resonances between $^{23}$Na$^{40}$K + $^{40}$K. These resonances have been attributed to long-range states of the triatomic complex in which the atoms and molecules retain their individual character~\cite{Wang2021}. One of these Feshbach resonances was recently used to form weakly-bound $^{23}$Na$^{40}$K$_2$ molecules by rf association~\cite{Yang2021}. Magnetic Feshbach resonances have also been observed for reactive $^{23}$Na$^{6}$Li$+^{23}$Na collisions, where the loss rate can be modulated by more than a factor of a hundred~\cite{Son2021}.

There remain many open questions in this field. First, we do not yet know the extent to which laser-induced loss via complexes is dominant. There are some systems in which there is no energetically allowed two-body pathway that leads to trap loss; at densities too low for three-body collisions, laser absorption seems to be the only available loss mechanism. However, for reactive systems and some systems involving excited atomic or molecular states, laser-free reactive or elastic processes may compete. Such losses may themselves occur either directly or via complexes. When complexes are involved, the factors that determine their lifetimes are poorly understood. Indeed, it is not even certain whether the complexes behave chaotically, with widths that are well described by RRKM theory. If they are chaotic, it is unclear whether the electron and nuclear spins are coupled into the chaotic bath, or are to some extent decoupled from it.

Motivated by these open questions, in this article we study collisional losses in an ultracold gas of $^{87}$Rb$^{133}$Cs molecules (hereafter RbCs), and in ultracold mixtures of RbCs+$^{87}$Rb and RbCs+$^{133}$Cs. We first examine RbCs+RbCs collisions. In section 1.1, we outline a model for loss of molecules from an optical trap via optical excitation and inelastic loss of two-molecule collision complexes. We show how the use of an intensity-modulated trap may be used to probe these loss mechanisms. In section 1.2, we present our method for creating an ultracold gas of RbCs molecules. In section 1.3, we show results comparing collisional loss of molecules from continuous-wave and modulated traps, for molecules prepared in three different hyperfine states. We observe the largest suppression of collisional losses in the modulated trap for molecules in the hyperfine ground state. We then proceed to study molecule loss in a reactive mixture of RbCs+Rb and a nonreactive mixture of RbCs+Cs in section 2. We begin by explaining our experimental method for measuring atom-molecule collision rates in our apparatus. In section 2.1, we show that the molecule loss rates for both the reactive and nonreactive collisions depend linearly on the atomic density; this confirms that the collisional loss in both mixtures has a two-body rate-limiting step. In section 2.2, we extract two-body rate coefficients from our measurements of molecule loss. We find rate coefficients that are lower than the limit of universal loss, but are similar for reactive and nonreactive collisions. Finally, in section 2.3, we compare loss from the nonreactive mixture of RbCs+Cs in intensity-modulated and CW traps. We find no significant change in the two-body rate coefficient when the optical trap is modulated and interpret this observation using our rate-equation model for the optical excitation of collision complexes.

\section{RbCs+RbCs Collisions}
\label{sec:MolMol}

\subsection{A rate-equation model for loss via bimolecular collision complexes}

We begin by examining molecule-molecule collisions in an optical trap, where the intensity of the light is modulated as a square-wave such that the molecules spend 75\% of each modulation cycle in the dark. In doing so, we partially suppress the optical excitation of two-body collision complexes, which is believed to be the dominant mechanism for collisional loss for pairs of molecules in the absolute ground state. When the trap light is off, complexes can form and break apart without the risk of destructive optical excitation. This leads to a reduction in the loss rate, with the maximum fractional reduction in loss simply equal to the duty cycle of the modulation. 

We model the rate of change of the densities of `free' molecules ${n}_\mathrm{m}$ and bimolecular complexes ${n}_\mathrm{c}$ with the rate equations 
\begin{equation}
    \dot{n}_\mathrm{m} = -k_{2}n_\mathrm{m}^{2} + \frac{2}{\tau_\mathrm{-1}}n_\mathrm{c}, \\
    \dot{n}_{c} = +\frac{1}{2}k_{2}n_\mathrm{m}^{2} - \frac{1}{\tau_\mathrm{-1}} n_\mathrm{c} - \frac{1}{\tau_\mathrm{inel}} n_\mathrm{c} - k_{\mathrm{laser}} I(t) n_\mathrm{c}.
\label{eq:RateEquations}
\end{equation}
This differs from previous treatments~\cite{Gregory2019, Bause2021, Gersema2021} in the inclusion of a term for inelastic loss of complexes.
Here $k_2$ describes the rate of formation of complexes and $k_\mathrm{laser}$ is the photon scattering rate of the complexes per unit intensity $I(t)$. The lifetime $\tau_\mathrm{-1}$ is the $1/e$ time for dissociation of the complexes back to free molecules in the initially prepared state, and $\tau_\mathrm{inel}$ describes the loss of molecules via a non-optical mechanism, such as conversion to a state other than the one in which the molecules are initially prepared. The lifetime of the complex in the dark is then $1/\tau_\mathrm{c}=1/\tau_{-1}+1/\tau_\mathrm{inel}$, and the probability that the molecules return to the initial state is $\Phi=\tau_\mathrm{inel}/(\tau_{-1}+\tau_\mathrm{inel})$.
We have previously measured $k_2$ by examining the rate of loss of molecules from a continuous-wave trap, which is rate-limited by the formation of complexes. The molecules have a temperature of $2$\,$\mu$K and $k_2=5.4\times10^{-11}$\,cm$^{3}$\,s$^{-1}$~\cite{Gregory2019, Gregory2020}, which is about a factor of two lower than the thermally averaged universal rate \cite{Frye2015}. In addition, we set $k_\mathrm{laser}=3\times10^{3}$\,W$^{-1}$\,cm$^{2}$\,s$^{-1}$ as measured in our previous experiments~\cite{Gregory2019} on molecules in their absolute ground state (where $\Phi=1)$. The behaviour of our model is however insensitive to this value provided that $k_\mathrm{laser}I(t) \gg 1/\tau_\mathrm{c}$ when the trap light is on, corresponding to the situation where the loss of complexes is strongly saturated.

\begin{figure}[t]
    \centering
        \includegraphics[width=0.75\textwidth]{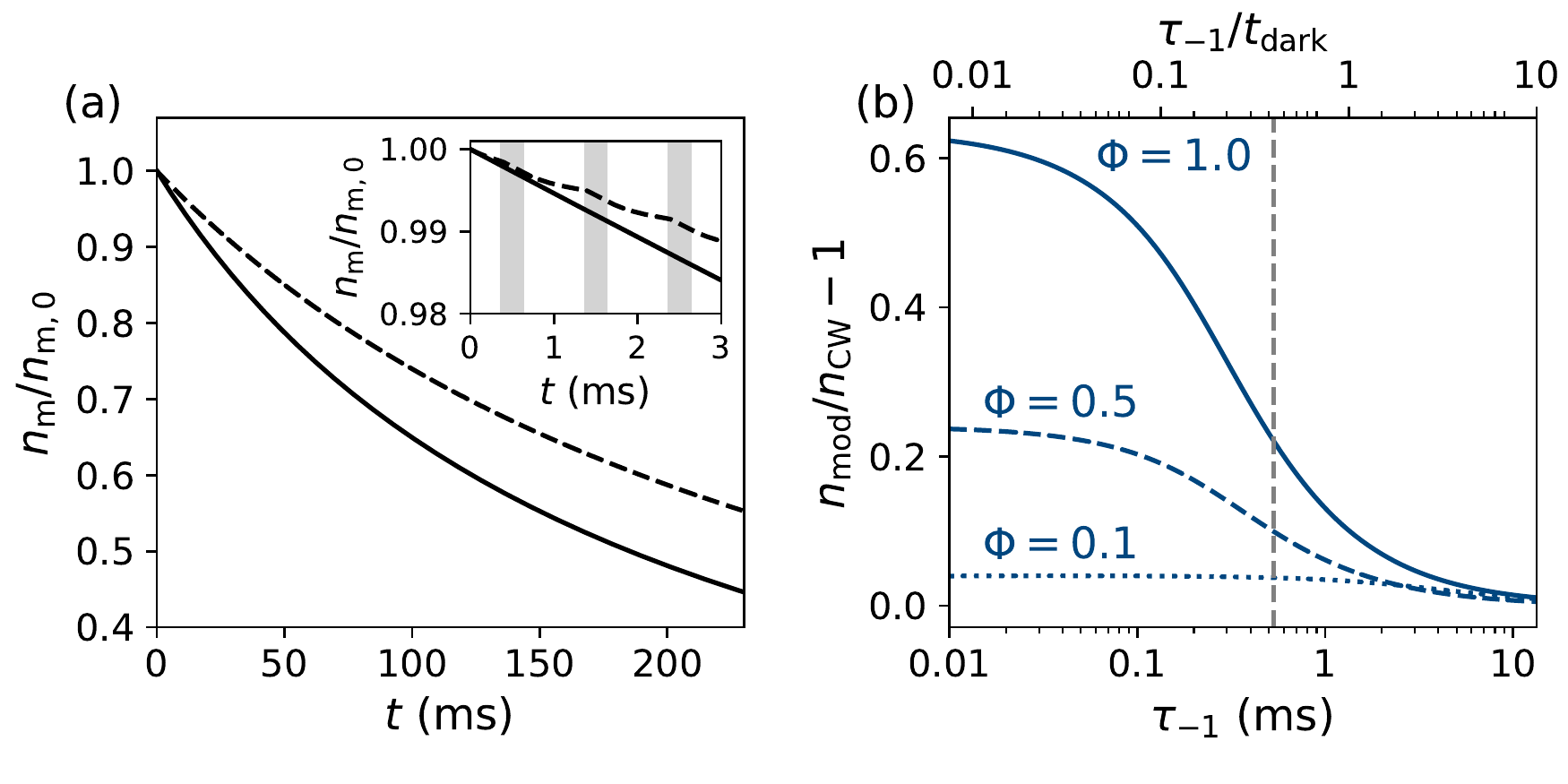}
    \caption{Suppression of loss of molecules using an intensity-modulated trap. We present results of the model described by Eq.~\ref{eq:RateEquations} for a trap modulation frequency of  $f_\mathrm{mod}=1$\,kHz and a $25\%$ duty cycle. The other parameters are set to the values determined in our previous work~\cite{Gregory2020}, namely  $k_2=5.4\times10^{-11}$\,cm$^{2}$\,s$^{-1}$, $k_\mathrm{laser}=3\times10^{3}$\,W$^{-1}$\,cm$^{2}$\,s$^{-1}$, $I=2\times10^4$\,W\,cm$^{-2}$, and $n_\mathrm{m,0}=10^{11}$\,cm$^{-3}$. (a)~Comparison between the variation in density of molecules $n_\mathrm{m}$ in a continuous-wave trap (solid line) and a modulated trap (dashed) for the case where $\Phi=1$ and $\tau_\mathrm{c}=\tau_\mathrm{-1}=0.53$\,ms. The inset highlights the change in density at short timescales. The shaded regions indicate time when the trap light is on. (b)~Effect of varying the decay time for complexes back to the initially prepared state $\tau_{-1}$ on the suppression of loss in the modulated trap. We plot the fractional difference between the density in CW ($n_\mathrm{CW}$) and modulated ($n_\mathrm{mod}$) traps after a 200\,ms hold time. The top horizontal axis expresses $\tau_{-1}$ as a fraction of the dark time in the trap $t_\mathrm{dark}$. The solid, dashed, and dotted lines shows the case for $\Phi=1$, 0.5, and 0.1, respectively. The vertical dashed line indicates the value $\tau_\mathrm{-1}$ used in (a).}
    \label{fig:Model}
\end{figure}

In Fig.~\ref{fig:Model}, we show the effect of modulating the trap intensity as a square wave by solving these rate equations for continuous-wave (CW) and modulated traps. For the modulated trap, the intensity is modulated at a frequency of 1\,kHz and the molecules spend 75\% of each cycle in the dark. In Figure~\ref{fig:Model}(a), we fix $\Phi=1$ and $\tau_\mathrm{c}=\tau_{-1}=0.53$\,ms, as previously measured for molecules in the spin-stretched hyperfine ground state~\cite{Gregory2020}. In this case we expect a slower rate of molecule loss in the modulated trap (dashed line) when compared to the CW trap (solid line). In Fig.~\ref{fig:Model}(b), we examine the effect of varying $\tau_\mathrm{-1}$ and $\Phi$, for a fixed hold time in the trap of 200\,ms. 
We find that significant suppression of the loss occurs only if $\tau_{-1}<\tau_\mathrm{inel}$ and $\tau_{-1}<t_\mathrm{dark}$, where $t_\mathrm{dark}$ is the duration the molecules spend in the dark during each cycle of the trap modulation.

The dark time can be varied in experiments by changing the frequency of the modulation, and we expect any suppression to become greater as the modulation frequency is reduced and the dark time increases correspondingly. Measurements in modulated traps thus allow one to distinguish optical loss from other loss mechanisms only if the lifetime of the complex is significantly shorter than the dark time and the optical loss dominates over any laser-free loss.

\subsection{Creating ultracold ground-state RbCs molecules}

For our experiments, we create a sample of molecules~\cite{McCarron2011, Koeppinger2014, Molony2014, Gregory2015, Molony2016} starting from an ultracold atomic mixture of Rb and Cs atoms confined to a magnetically levitated crossed optical dipole trap (with a wavelength, $\lambda=1550$\,nm)~\cite{McCarron2011}. We first form weakly bound molecules from the atomic mixture using magnetoassociation on an interspecies Feshbach resonance at 197\,G~\cite{Koeppinger2014}. Following this, we remove the remaining atoms from the trap using the Stern-Gerlach effect~\cite{Koeppinger2014}. We then transfer the molecules into an optical trap where the intensities of the beams are modulated as a square wave. This is achieved by ramping up the intensity of the modulated trap and switching off the 1550\,nm CW trap and magnetic levitation gradient.

The modulated trap is formed from a single beam $(\lambda=1064$\,nm) in a bow-tie configuration. The trap frequencies experienced by molecules in this trap are $[\omega_x, \omega_y, \omega_z]/2\pi=[96(2), 160(3), 185(3)]$\,Hz. The square-wave intensity modulation is achieved by blocking the light using an optical chopper wheel. Using this method, we can modulate the trap intensity with cycle frequencies of up to 5\,kHz. We find significant trap loss for modulation frequencies below $1$\,kHz, consistent with loss due to parametric heating resonances, as occurs in time-dependent potentials that are not fully in the time-averaged trap regime~\cite{Schnelle2008, Henderson2009, ROberts2014}.

We transfer the molecules in the modulated trap to the $^{1}\Sigma$ rovibrational ground state using STImulated Raman Adiabatic Passage (STIRAP)~\cite{Molony2014, Gregory2015, Molony2016}. The STIRAP is performed at a magnetic field of $B=181.6$\,G during the trap dark time. Immediately after STIRAP, the sample typically consists of up to 5000 molecules with a mean density of~$\sim10^{11}$\,cm$^{-3}$. STIRAP prepares the molecules in a single hyperfine level of the rovibrational ground state.

For RbCs at $B=181.6$\,G the rovibrational ground state consists of 32 hyperfine states spread across 1.3~MHz with separations between neighbouring states ranging from 10 to 100~kHz as shown in Fig.~\ref{fig:ModStates}(a). We label these hyperfine states by $(n, m_{f,\mathrm{RbCs}})_k$ where $n$ is the quantum number for rotational angular momentum, $m_F=m_\mathrm{Rb}+m_\mathrm{Cs}+m_n$ is the sum of the angular momentum projections for the nuclear spins $m_\mathrm{Rb}, m_\mathrm{Cs}$ and the rotation of the molecule $m_n$, and $k$ is an index that counts up the states in order of increasing energy for a given value of $n$ and $m_{f,\mathrm{RbCs}}$. For magnetic fields above $90$\,G the spin-stretched state $(0,5)_0$ state is the hyperfine ground state of RbCs. With the STIRAP, we are able to populate either the $(0,5)_0$ or $(0,4)_1$ hyperfine states directly. In this work, we also prepare molecules in $(0,4)_0$, which we achieve using a pair of one-photon $\pi$-pulses to drive transitions in the molecule coherently between $n=0$ and $n=1$~\cite{Gregory2016}. The states used in this work are highlighted in Fig.~\ref{fig:ModStates}(a).

\subsection{RbCs+RbCs in an intensity-modulated trap}

Our understanding of the formation and optical excitation of molecule-molecule collision complexes is far from complete. This is evidenced by observations in $^{23}$Na$^{39}$K, $^{23}$Na$^{40}$K, and $^{23}$Na$^{87}$Rb~\cite{Bause2021, Gersema2021} where no suppression of loss in modulated traps was seen, despite RRKM predictions of lifetimes for complexes~\cite{Christianen2019DOS} that are much shorter than the dark time in the trap. We have previously observed a suppression of loss for RbCs molecules in the spin-stretched hyperfine ground state, so here we explore collisions of  molecules prepared in different hyperfine states, where other loss channels and mechanisms could affect the lifetime of the complex.

To measure the effect of the dark time in the modulated trap, we measure the number of molecules remaining after a 200\,ms hold in the modulated trap, both with ($N_\mathrm{mod+CW}$) and without ($N_\mathrm{mod}$) an additional CW source of 1550\,nm light. The CW light is derived from the 1550\,nm trap used to prepare the Feshbach molecules, and has total peak intensity of $\sim3\times10^{2}$\,W\,cm$^{-2}$. This intensity is sufficiently low that it does not significantly affect the trap frequencies or the trap depth experienced by the molecules, but high enough to remove complexes continuously from the trap~\cite{Gregory2020}. By performing a comparative measurement, 
we remove any sensitivity to the effects of residual heating and associated evaporative loss that may arise due to the modulation of the trap intensity. To measure the molecule number, we reverse the STIRAP and association process, to break the molecules back apart into their constituent atoms for detection by absorption imaging. For each measurement of loss, we perform 50 interleaved measurements of $N_\mathrm{mod}$ and $N_\mathrm{mod+CW}$ and extract a mean and standard error from the resulting distributions.

\begin{figure}[t]
    \centering
        \includegraphics[width=0.7\textwidth]{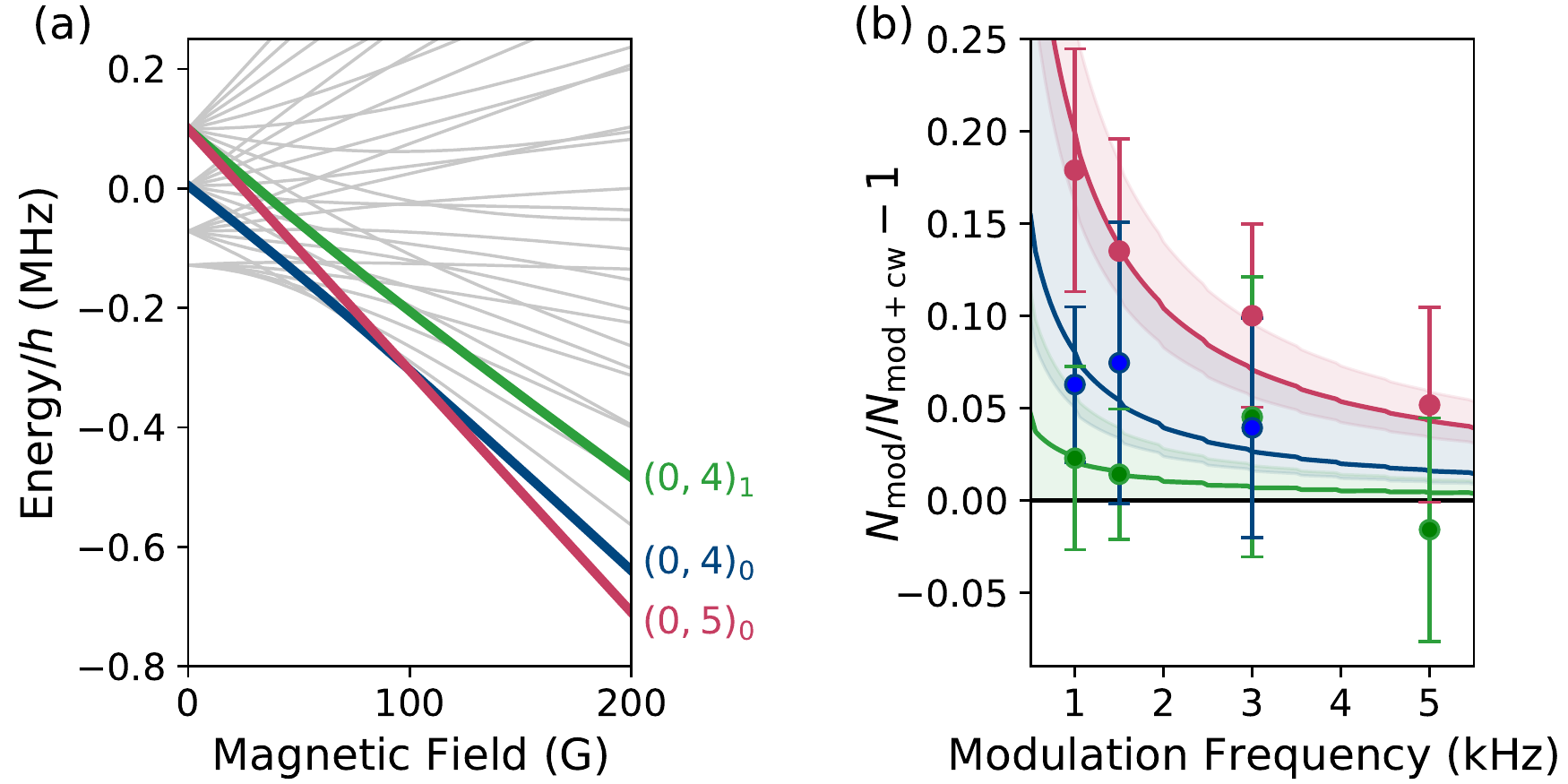}
    \caption{(a)~Hyperfine Zeeman structure of RbCs in the rovibrational ground state. The states used in this work are highlighted in bold colours and labelled by $(n, m_{f,\mathrm{RbCs}})_k$ as defined in the main text. (b)~Fractional change in the molecule number $N_\mathrm{mod}/N_\mathrm{mod+CW}-1$ as a function of the modulation frequency of the trap light intensity. Results are shown for molecules in $(0,5)_0$ in red, $(0,4)_0$ in blue, and $(0,4)_1$ in green, with all measurements performed at a magnetic field of 181.6\,G. We observe a smaller suppression of the collisional loss for molecules not occupying the hyperfine ground state $(0,5)_0$. The lines show fits to the results using the numerical model in Eq.~\ref{eq:RateEquations} for $\Phi=1$, with the 1$\sigma$ uncertainty in the lifetime of the complex in the dark indicated by the shaded regions for each fit. }
    \label{fig:ModStates}
\end{figure}

We characterise the suppression of loss due to the dark time by calculating the fractional difference in the molecule number $N_\mathrm{mod}/N_\mathrm{mod+CW}-1$. This is shown as a function of modulation frequency in Fig.~\ref{fig:ModStates}(b) for molecules prepared in several different hyperfine states. For molecules in $(0,5)_0$ we observe a suppression of loss characterised by $N_\mathrm{mod}/N_\mathrm{mod+CW}-1>0$. The suppression is greatest at the lowest modulation frequencies we are able to reach where $t_{\rm{dark}}/\tau_{\rm{-1}}$ is greatest, see Fig.~\ref{fig:Model}(b)). We fit the results using the rate equation model (Eq.~\ref{eq:RateEquations}), making the assumption that the molecules remain at a fixed temperature. We fix $\Phi=1$ because the $(0,5)_0$ state is both spin-stretched and is the lowest-energy hyperfine state; the molecules do not have sufficient kinetic energy to leave the complex in any other state. As such, the lifetime of the collision complex in the dark is the only free parameter in the fitting. We find an optimal value for the lifetime of the (RbCs)$_2$ collision complex is 0.8(3)\,ms in the dark, where the number in brackets is the 1$\sigma$ uncertainty.
This is consistent with our previously measured value of 0.53(6)\,ms~\cite{Gregory2020}.

Fig.~\ref{fig:ModStates}(b) also shows similar measurements for molecules prepared in the higher-energy hyperfine states $(0,4)_0$ and $(0,4)_1$. In these cases we observe a lower suppression of the loss. We first analyse results with the assumption $\tau_\mathrm{inel}\rightarrow\infty, \Phi=1$. In this limit, our results would suggest that the lifetime of the complex in the dark depends on the hyperfine state, which might be associated with an increase in the effective density of states. This change might be caused by the increased number of nuclear spin arrangements that have the same value of the total spin projection $M_F=m_{f,1}+m_{f,2}$ as the incoming pair. For the $(0,4)_0$ state our results are consistent with $\tau_\mathrm{c}=\tau_{-1}=2.1(1.3)$\,ms. In contrast, for the $(0,4)_1$ state our results are consistent with no suppression of loss and we can place only a lower limit on the lifetime of the complex in the dark, $\tau_\mathrm{c}=\tau_{-1}>3.3$\,ms (at the 68\% confidence level).

An alternative interpretation of the reduced suppression for the states with $m_{f,\mathrm{RbCs}}=4$ is the presence of one or more additional loss channels. Collisions between molecules in either state have at least one energetically accessible inelastic channel that conserves $M_F$. 
Decay of complexes to form molecules in different states would be observed as loss in our experiments, because we detect molecules only in the specific hyperfine state in which they are prepared. To test this interpretation, we restrict  $\tau_{-1}$ to the range of values in the 68\% confidence interval found for the $(0,5)_0$ state, $\tau_\mathrm{-1}=0.8(3)$\,ms, and fit the results for $(0,4)_0$ and $(0,4)_1$ with $\tau_\mathrm{inel}$ as a free parameter. 

For molecules prepared in $(0,4)_0$ there is one energetically available channel for spin exchange, with the molecules exiting the complex in $(0,5)_0+(0,3)_0$. This combination of states is higher in energy than the prepared state by only $k_\mathrm{B}\times0.16$\,$\mu$K, which is much smaller than the temperature of the molecules $T=2\,\mu$K. For molecules in $(0,4)_0$, our results find an optimum value $\tau_\mathrm{inel}=0.2^{+0.5}_{-0.1}$, leading to $\Phi=0.2^{+0.4}_{-0.1}$. For molecules in $(0,4)_1$, several spin exchange channels are available. In addition, there is the possibility of exchange between nuclear spins in the same molecule such that it exits in the lower-energy $(0,4)_0$ state. For the $(0,4)_1$ state, we find $\tau_\mathrm{inel}<0.13$\,ms and $\Phi<0.08$ at the 68\% confidence level.

This analysis demonstrates that our experimental results for excited molecular states can be explained either by a longer lifetime $\tau_{-1}$  or by the presence of laser-free inelastic decay. We cannot at present distinguish between these explanations.

\section{Collisions of RbCs with Rb or Cs}
\label{sec:AtmMol}

We now examine collisions between RbCs molecules and the constituents of the atomic mixture in which they are prepared. To achieve this, the Rb-Cs mixture is prepared in a purely optical potential ($\lambda=1550$\,nm), with no magnetic levitation to support against gravity. We then perform magnetoassociation on the same Feshbach resonance as before ($B_0= 197$\,G), but the transfer to the $(0,5)_0$ ground state instead takes place 1\,ms after magnetoassociation, while the atoms are still present in the trap. The 1\,ms hold is necessary to allow the magnetic field to reach $181.6$\,G and become sufficiently stable for efficient STIRAP. With the molecules in the ground state, one atomic species is removed by ramping the magnetic field to 21.3\,G and turning on the appropriate repump and cooling light from the magneto-optical trap for 3\,ms. After the unwanted atoms are removed, we either hold the atom-molecule mixture in the optical trap at 21.3\,G, or ramp the magnetic field to 181.6\,G over a further 2\,ms for the lifetime measurement. At the end of the measurement, prior to dissociation and imaging, the magnetic field is jumped back to 21.3~G to remove the atomic species left after the first removal pulse. We also measure molecular lifetimes without atoms using this sequence by removing both species of atoms in the first removal step. This provides a useful consistency check with the measurements discussed in section~\ref{sec:MolMol} to verify the efficacy of the atom-removal procedure. We also use such measurements to account for background loss of molecules, as described later.

Our experimental sequence produces up to $\sim8\times10^{5}$ ($3\times10^{5}$) Rb (Cs) atoms at a temperature of 1.1(1)\,$\upmu$K, and mean density of typically $\sim10^{12}$\,cm$^{3}$\,s$^{-1}$. Each species is prepared in the hyperfine ground state $(f_\mathrm{Rb}=1,m_{f,\mathrm{Rb}}=1)$ for Rb and $(f_\mathrm{Cs}=3,m_{f,\mathrm{Cs}}=3)$ for Cs. The temperature and density of the atoms remain constant over the duration of the experiments (typical timescales of $\sim$10 ms). The molecules on the other hand have a temperature of $1.3(3)$\,$\upmu$K, and and mean starting density $\sim10^{10}$\,cm$^{-3}$. The reduced starting density for the molecules in these measurements is caused by a combination of lower trap frequencies and loss of molecules from inelastic collisions between atoms and Feshbach molecules during the association procedure.

\begin{figure}[t]
    \centering
        \includegraphics[width=0.5\textwidth]{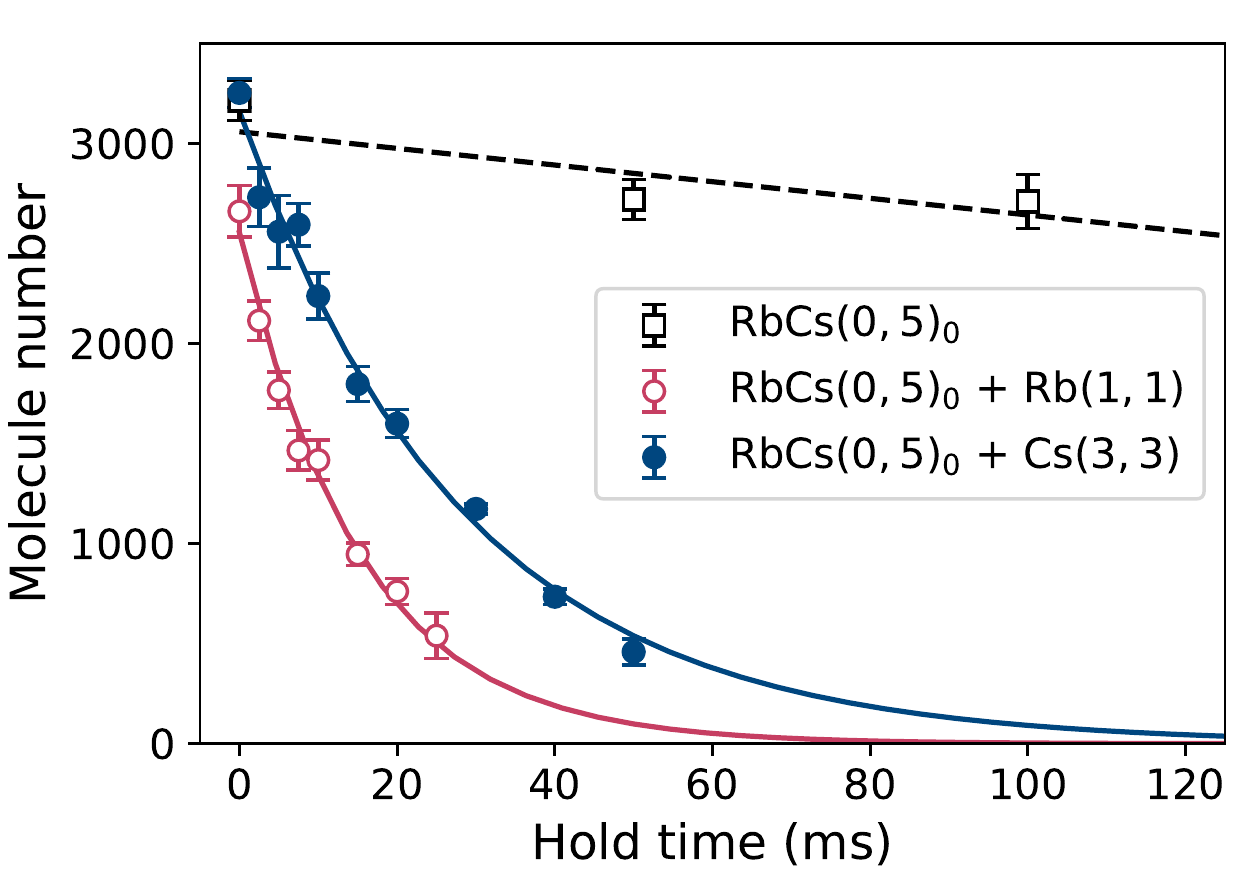}
    \caption{Example molecule loss measurements performed at 181.6\,G for molecules alone and for mixtures of RbCs+Rb and RbCs+Cs. The molecules are prepared in the $(n=0, m_{f,\mathrm{RbCs}}=5)_0$ state which is the hyperfine ground state of the molecule at this magnetic field. Collisional loss of RbCs molecules alone is shown by the black squares with a linear fit to the results indicated by the dashed line. Empty red circles show loss of RbCs molecules when prepared with ground-state Rb atoms $(f_\mathrm{Rb}=1, m_{f,\mathrm{Rb}}=1)$, such that the loss is dominated by reactive RbCs+Rb collisions. Filled blue circles show loss of RbCs molecules with ground-state Cs atoms $(f_\mathrm{Cs}=3, m_{f,\mathrm{Cs}}=3)$, where loss is dominated by nonreactive RbCs+Cs collisions. The solid lines are exponential fits incorporating the molecule-only loss following Eq.~\ref{eq:ExpFit} as explained in the main text.  }
    \label{fig:ExampleLoss}
\end{figure}

The lifetime of RbCs molecules in the presence of each species of atom is shown in Fig.~\ref{fig:ExampleLoss}. To account for the background variation in the number of molecules due to molecule-molecule collisions, we also measure a lifetime for molecules alone. As the atom-molecule loss is much faster than the molecule-only loss in all measurements presented, we fit the molecule-only loss with a straight line with starting number $N_0$ and gradient $m$ as shown. We then fit the variation in the number of molecules $N_\mathrm{m}$ for each atom-molecule combination with an exponential function,
\begin{equation}
N_\mathrm{m}(t) =  N_0 \left(1-mt\right) \exp{\left(\frac{-t}{\tau}\right)}.
\label{eq:ExpFit}
\end{equation}
Here, $t$ is the hold time in the trap, $\tau$ is the $1/e$ lifetime for the atom-molecule collisions, $N_0$ is the initial number of molecules, and the first term in brackets represents the normalisation to the molecule-only background. For the curves shown in Fig.~\ref{fig:ExampleLoss}, we extract $1/e$ lifetimes of 16(1)\,ms for reactive RbCs$+$Rb collisions and 30(2)\,ms for nonreactive RbCs$+$Cs collisons. However, as the densities of Rb and Cs are different, direct comparison of these time constants is not immediately useful; we present density-normalised rate coefficients, which may be properly compared, in section 2.3.

\subsection{Density dependence}

Studying the change in molecule loss rate as a function of atom density can yield insight into the kinetics of the underlying loss mechanism. For reactive RbCs+Rb collisions, the energetically allowed atom-exchange reaction is likely to be the dominant loss mechanism, although it is possible that laser-induced loss may also be important \cite{Nichols2021}. As this is a two-body (atom+molecule) process, we expect that the rate of loss of molecules will depend linearly upon the density of atoms. For nonreactive RbCs+Cs collisions however, the expectation is less clear. Possible mechanisms for loss in the nonreactive mixture include (but may not be limited to) optical excitation of two-body (atom-molecule) complexes and three-body (atom-atom-molecule) collisions. We expect the loss rate associated with these processes to depend on the atom density, or the square of the atom density, respectively.  

\begin{figure}[t]
    \centering
        \includegraphics[width=0.7\textwidth]{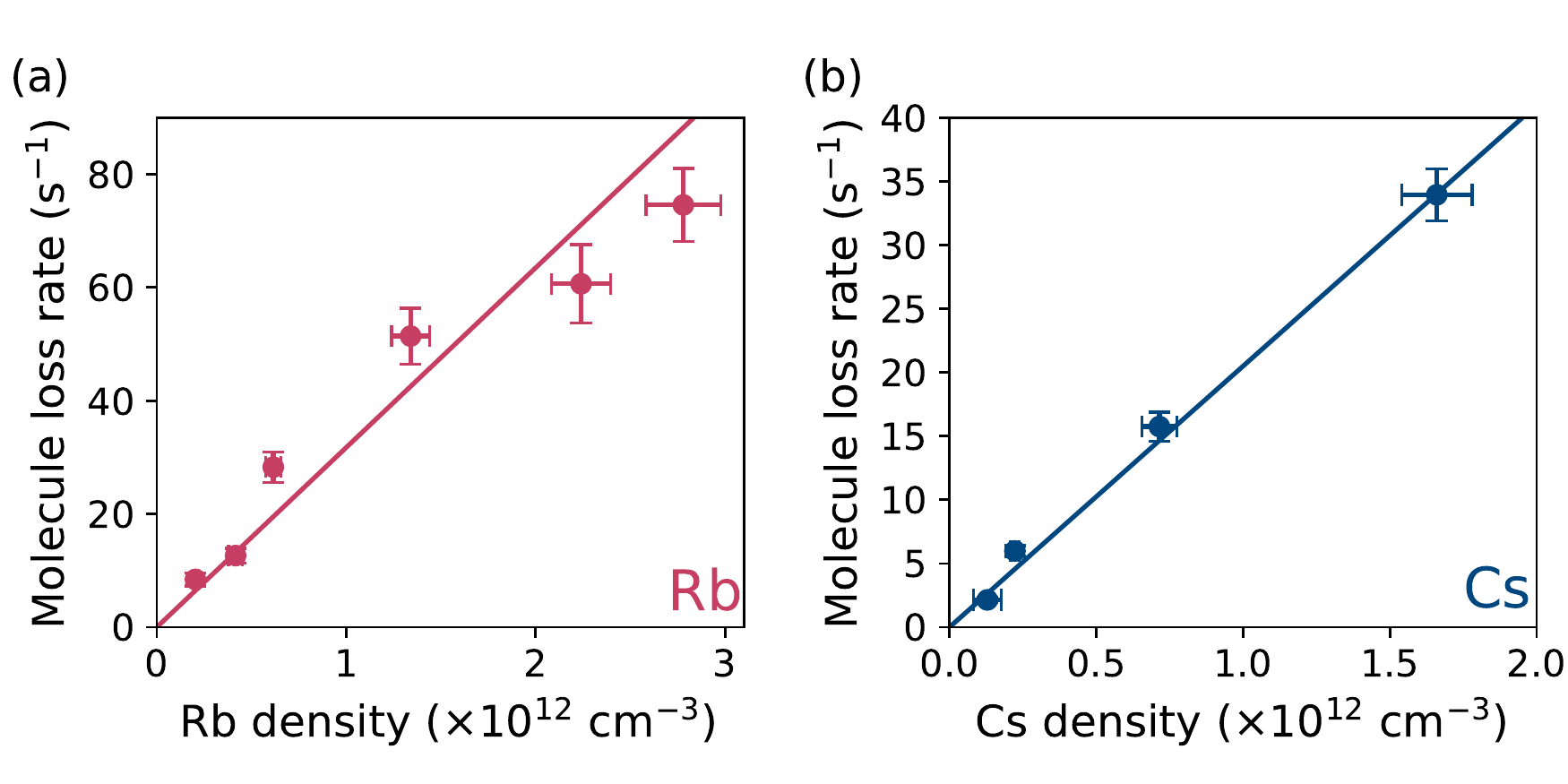}
    \caption{Dependence of the molecule loss rate on the mean atomic density. We control the atomic density by pulsing on near-resonant light that reduces the number of atoms remaining in the trap. (a) Loss rate of RbCs molecules in $(n=0,m_{f,\mathrm{RbCs}}=4)_1$ as a function of the Rb density. The solid line is a linear fit to the results constrained to go through the origin with gradient $3.2(2)\times10^{11}$\,cm$^{3}$\,s$^{-1}$. (b) Loss rate of RbCs molecules in $(0,5)_0$ as a function of the Cs density. The solid line is a linear fit to the results constrained to go through the origin with gradient $2.05(7)\times10^{11}$\,cm$^{3}$\,s$^{-1}$. } 
    \label{fig:DensityDependence}
\end{figure}

To vary the atom density we use resonant light to remove a fraction of the chosen atomic species from the trap. This is performed in parallel with the removal of the unwanted atomic species after STIRAP. The additional light is tuned to be resonant with the $5^{2}\mathrm{S}_{1/2}(f_\mathrm{Rb}=1,m_{f,\mathrm{Rb}}=1)\rightarrow5^{2}\mathrm{P}_{3/2}(2,2)
$ electronic transition for Rb and the $6^{2}\mathrm{S}_{1/2}(f_\mathrm{Cs}=3,m_{f,\mathrm{Cs}}=3)\rightarrow6^{2}\mathrm{P}_{3/2}(4,4)
$ for Cs. The light is horizontally polarised and delivered in a collimated beam of radius $\sim1$\,mm that propagates orthogonal (designated $x$ axis) to the 21.3\,G magnetic field (along the $z$ axis). Throughout the resonant light pulse, we also switch on cooling light from the magneto-optical trap which removes any atoms that spontaneously decay into the $f_\mathrm{Rb}=2$ state of Rb and the $f_\mathrm{Cs}=4$ state of Cs. We observe loss of atoms with an exponential time constant of 88(2)\,$\upmu$s for Rb and 6.4(2)\,$\upmu$s for Cs. The difference in removal rate is primarily due to the different laser intensities used for the different species. Using this method, we do not observe any significant build-up of population in other atomic hyperfine states, such that the atoms that survive the resonant light pulse remain in the target $(1, 1)$ state for Rb and $(3,3)$ state for Cs. However, we do observe that the temperature of the atoms increases with the duration of the light pulse, with approximately a factor of 2 greater increase in the direction of the laser propagation. This change in temperature is taken into account when calculating the atomic density.

We present measurements of molecule loss rate as a function of mean atom density in Fig.~\ref{fig:DensityDependence}(a,b). In Fig.~\ref{fig:DensityDependence}(a), we examine the reactive combination RbCs+Rb. Specifically, we measure collisions between molecules prepared in $(0,4)_1$ with ground-state Rb atoms. As expected, we observe a linear dependence of the molecule loss rate on atom density. A linear fit to the results, constrained to pass through the origin, yields a gradient of $3.3(2)\times10^{11}$\,cm$^{3}$\,s$^{-1}$. In Fig.~\ref{fig:DensityDependence}(b), we show a measurement using the nonreactive combination of RbCs+Cs. In this case the molecules are prepared in $(0,5)_0$ so that both the Cs atoms and RbCs molecules occupy their respective hyperfine ground states. For the nonreactive combination we also find a linear dependence of the molecule loss rate on the atom density, this time with a gradient of $2.05(7)\times10^{11}$\,cm$^{3}$\,s$^{-1}$. This indicates that the loss mechanisms for both reactive and nonreactive collisions have rate-limiting steps that depends on a two-body RbCs+atom collision.

\subsection{Second-order rate coefficients for collisions of RbCs with Rb or Cs}
\label{sec:RateCoefficient}

\begin{figure}[t]
    \centering
        \includegraphics[width=0.7\textwidth]{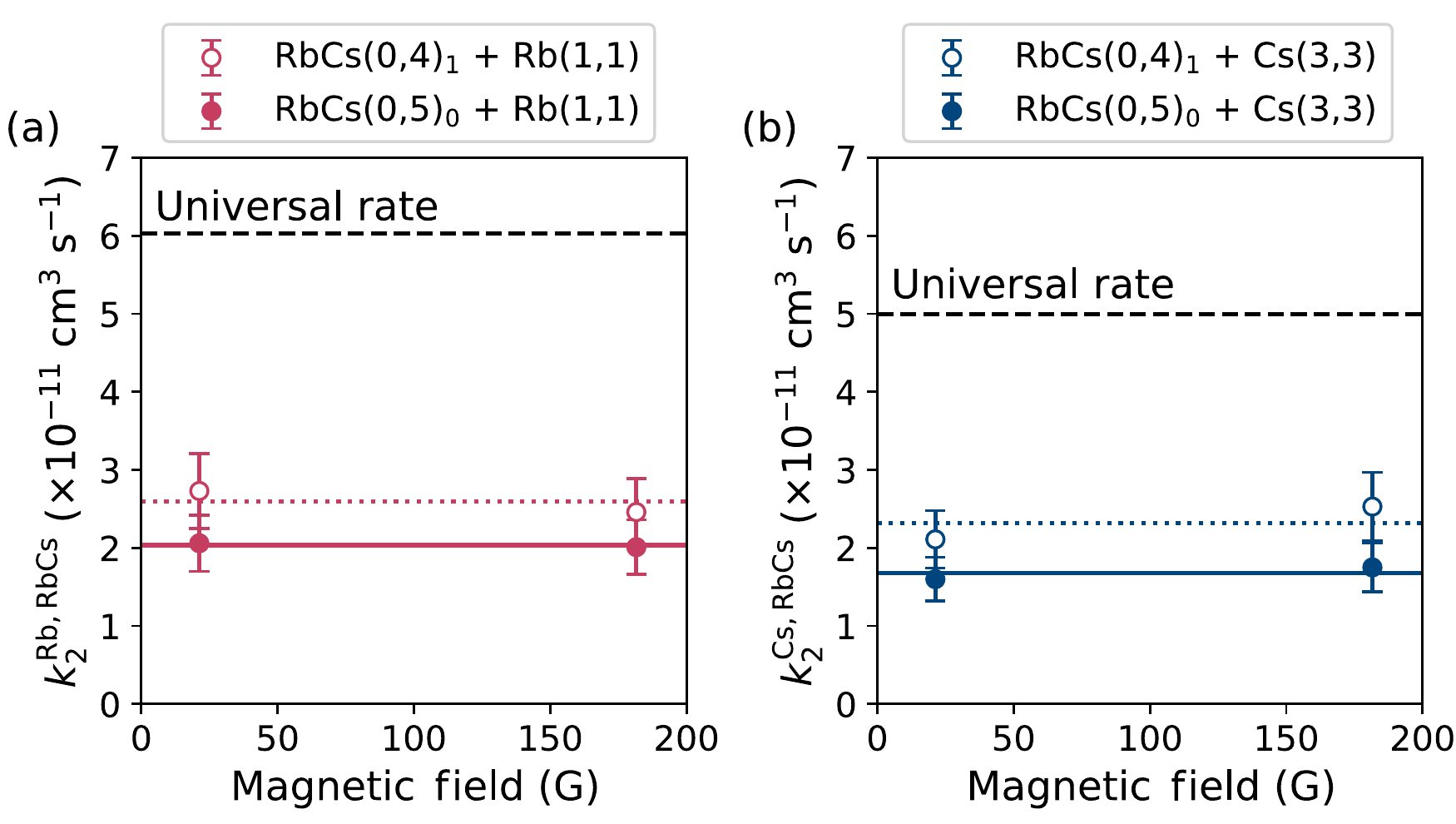}
    \caption{Second-order loss rate coefficients $k^\mathrm{a,m}_2$ measured for RbCs molecules collisions with ground-state (a) Rb (b) Cs atoms. Measurements were performed at magnetic fields of 21.3\,G and 181.6\,G. Filled and empty circles show the result of measurements performed with RbCs molecules in the $(n=0,m_{f,\mathrm{RbCs}}=5)_0$ and $(0,4)_1$ states respectively. The coloured solid horizontal line indicates the average of the two loss rates measured for molecules in the $(0,5)_0$ state, and the dashed line is the average for molecules in the $(0,4)_1$ state. In all measurements the atoms occupy their hyperfine ground states, either ($f_\mathrm{Rb}=1, m_{f,\mathrm{Rb}}=1$) or ($f_\mathrm{Cs}=3, m_{f,\mathrm{Cs}}=3$) for Rb and Cs respectively. The black dashed line indicates the thermally averaged universal rate for the particular atom-molecule combination.}
    \label{fig:Bdependence}
\end{figure}

As both reactive and nonreactive atom-molecule collisions appear to be rate-limited by second-order kinetics, we can quantify the loss rate using a second-order rate coefficient $k_2^\mathrm{a,m}$. 
To extract a second-order rate coefficient from a measurement of molecule loss, we model the loss of molecules due to atom-molecule collisions with the rate equation 
\begin{equation}
    \dot{n}_\mathrm{m}(\textbf{r}, t) = -k^\mathrm{a,m}_2 n_\mathrm{a}(\textbf{r}, t) n_\mathrm{m}(\textbf{r}, t),
\label{eq:rate-am}
\end{equation}
where $n_\mathrm{a}, n_\mathrm{m}$ represent the densities of atoms and molecules, respectively. The rate of change of the number of molecules can be obtained by integrating Eq.\ \ref{eq:rate-am}, giving
\begin{equation}
\dot{N}_\mathrm{m}(t)=-k^\mathrm{a,m}_2 \int{n_\mathrm{a}(\textbf{r}, t) n_\mathrm{m}(\textbf{r}, t)}{d^{3}\textbf{r}}.
\label{eq:RateMol}
\end{equation}
The term in the integral contains the overlap between the distributions of atoms and molecules, and we can use this integral to define a mean interspecies density 
\begin{equation}
     \bar{n}_\mathrm{a,m} = \frac{1}{N_\mathrm{m}(t)}\int{n_\mathrm{a}(\textbf{r}, t) n_\mathrm{m}(\textbf{r}, t)}{d^{3}\textbf{r}}.
\end{equation}
If we assume the atoms and molecules are thermally distributed in the harmonic region of the trap with temperatures $T_\mathrm{a}$ and $T_\mathrm{m}$, respectively, then,
\begin{equation}
    \bar{n}_\mathrm{a,m} = N_\mathrm{a} F_z(\Delta_z) \left[ \frac{m_\mathrm{m}\omega_\mathrm{m}^2m_\mathrm{a}\omega_\mathrm{a}^2}{2\pi k_\mathrm{B}(m_\mathrm{m}\omega_\mathrm{m}^2T_\mathrm{a} + m_\mathrm{a}\omega_\mathrm{a}^2T_\mathrm{m})} \right]^{3/2},
\label{eq:InterspeciesDensity}
\end{equation}
where $N_\mathrm{a}$ is the number of atoms (which remains constant over the duration of the measurement), $k_\mathrm{B}$ is the Boltzmann constant, $m_\mathrm{a}$, $m_\mathrm{m}$ are the masses and $\omega_\mathrm{a}, \omega_\mathrm{m}$ are the geometrically-averaged trap frequencies experienced by the atoms and molecules, respectively. $F_z(\Delta_z)$ describes the reduction in overlap due to the difference in gravitational sag $\Delta_z$ between the two clouds \cite{Guttridge2017},
\begin{equation}
    F_z(\Delta_z)= \exp\left[-\frac{m_\mathrm{m}\omega_\mathrm{z,m}^{2}m_\mathrm{a}\omega_\mathrm{z,a}^{2}\Delta_z^2}{2k_\mathrm{B}(m_\mathrm{a}T_\mathrm{m}\omega_\mathrm{z,a}^{2} + m_\mathrm{m}T_\mathrm{a}\omega_\mathrm{z,m}^{2})}\right],
    \label{eq:sagcompensation}
\end{equation}
where $\omega_\mathrm{z,a}$, $\omega_\mathrm{z,m}$ are the vertical trap frequencies of the atoms and molecules, respectively. The largest difference in gravitational sag in our experiments, \mbox{$\Delta_z=2.7$\,$\upmu$m}, is between Rb and RbCs. In this case, we find $F_z(2.7\,\upmu\mathrm{m})=0.98$; the difference in gravitational sag changes the interspecies density by only 2\%, which is much less than the typical uncertainty in either the atom or the molecule density alone~(each typically $\sim10\%$). We therefore neglect the $F_z(\Delta_z)$ term in Eq.~\ref{eq:InterspeciesDensity}. 
Returning to Eq.~\ref{eq:RateMol}, we have
\begin{equation}
    \dot{N}_\mathrm{m}(t)=-k^\mathrm{a,m}_2 \bar{n}_\mathrm{a,m} N_\mathrm{m}.
\end{equation}
We have confirmed experimentally that the atom number and temperature do not change appreciably over the course of the measurements presented. If we additionally assume that the molecule temperature does not change significantly, then $\bar{n}_\mathrm{a,m}$ is a constant. This produces pseudo-first-order kinetics with the solution
\begin{equation}
N_\mathrm{m}(t) = N_0 \exp{\left(-k^\mathrm{a,m}_2 \bar{n}_\mathrm{a,m} t\right)},
\label{eq:AtomMolLoss}
\end{equation}
where $N_0$ is the initial number of molecules.  This allows us to extract two-body rate coefficients from the measured exponential time constants.

We first compare the loss rates measured with molecules prepared in the hyperfine ground state $(0,5)_0$ at 181.6\,G. Example molecule loss measurements are presented in Fig.~\ref{fig:ExampleLoss}. For the reactive combination RbCs+Rb , we find $k_2^\mathrm{a,m}=2.0(4)\times10^{-11}$\,cm$^{3}$\,s$^{-1}$, while for the nonreactive combination RbCs+Cs we find $k_2^\mathrm{a,m}=1.8(3)\times10^{-11}$\,cm$^{3}$\,s$^{-1}$. Additional loss rates measured at a magnetic field of 21.3\,G and for molecules prepared in $(0,4)_1$ are presented in Fig.~\ref{fig:Bdependence}. We observe no significant variation with the large change in magnetic field, and only marginally higher loss rates for atom-molecule collisions involving molecules prepared in $(0,4)_1$. 

\begin{figure}[t]
    \centering
        \includegraphics[width=0.45\textwidth]{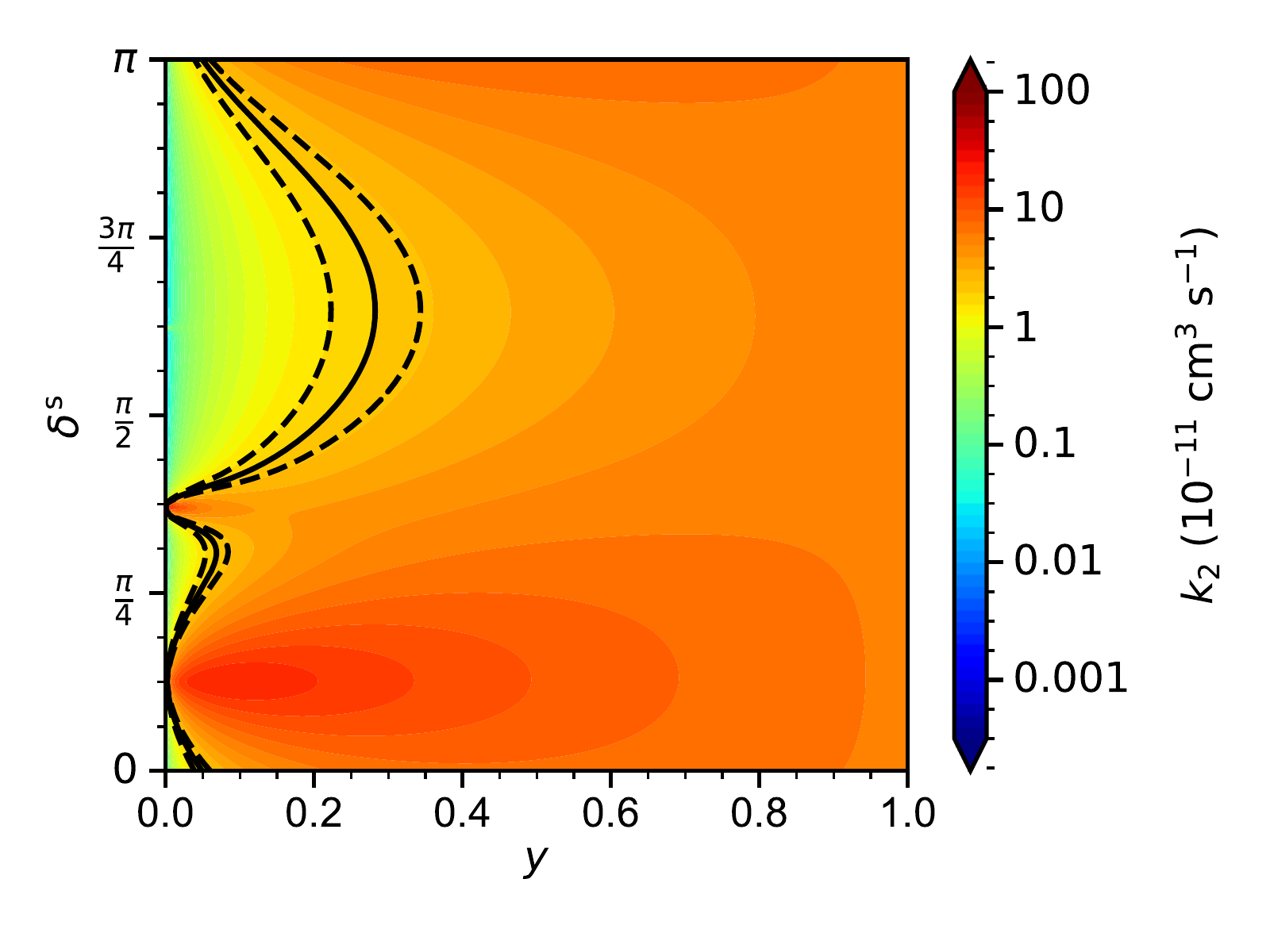}        \includegraphics[width=0.45\textwidth]{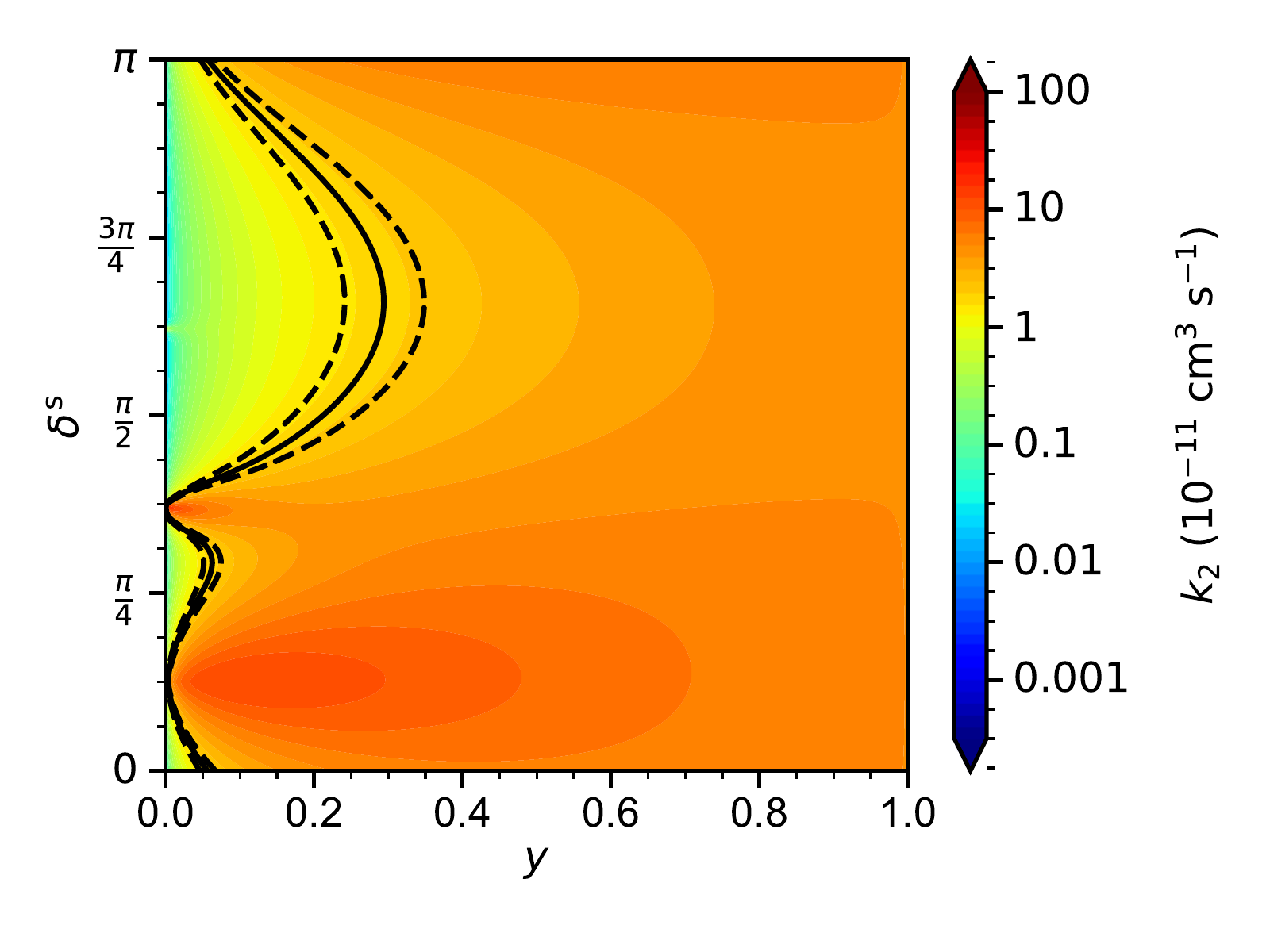}
    \caption{Thermally averaged two-body loss rate coefficient $k_2$ from the QDT model, as a function of loss parameter $y$ and short-range phase shift $\delta^\mathrm{s}$ for RbCs+Rb (left) and RbCs+Cs (right). Solid and dashed black lines show the experimental value and uncertainty. \label{fig:QDT}
    }
\end{figure}

We compare the loss rates measured in experiments to a single-channel model based on quantum defect theory (QDT) \cite{Idziaszek2010, Frye2015}. The model and its underlying theory have been described at length elsewhere \cite{Gao:C6:1998,Gao:2008,Frye2015}, so we omit further description here; we have previously applied it to RbCs+RbCs and Rb+CaF collisions \cite{Gregory2019,Jurgilas2021,Jurgilas2021b}. The long-range interactions are approximated by their leading term $-C_6R^{-6}$ and the short-range interactions are modelled by an absorbing boundary condition. We use values of $C_6$ from \.Zuchowski \emph{et al.} \cite{Zuchowski:2013}. The boundary condition is parameterized by the loss parameter $0\leq y\leq 1$ of Idziaszek and Julienne \cite{Idziaszek2010} and a short-range phase shift $\delta^\mathrm{s}$, which is related to the scattering length and controls interference effects including resonances. In the limit of $y=1$ all flux that is transmitted past the long-range potential is lost; this is termed the universal limit. The thermally averaged universal loss rates at the current experimental temperatures are $6.0\times10^{-11}$\,cm$^{3}$\,s$^{-1}$ for RbCs+Rb and $5.0\times10^{-11}$\,cm$^{3}$\,s$^{-1}$ for RbCs+Cs; these are indicated by horizontal dashed lines in Fig.~\ref{fig:Bdependence}.

The results of the QDT model as a function of $y$ and $\delta^\mathrm{s}$, for RbCs+Rb and RbCs+Cs, are shown in Fig.~\ref{fig:QDT}, with the experimental results and their uncertainties. The present results limit the loss parameter to less than 0.35, with the most likely range being $0.2<y<0.3$ for both species. It is notable that the ranges are very similar for the two systems, even though one is potentially reactive and the other is non-reactive. The ranges also overlap with the result for RbCs+RbCs, $y=0.26(3)$ \cite{Gregory2019}. All three results are consistent with a recent prediction \cite{Christianen2021} that rapid loss from collision complexes would result in an effective loss rate described by $y=0.25$.

\begin{figure}[t]
    \centering
        \includegraphics[width=\textwidth]{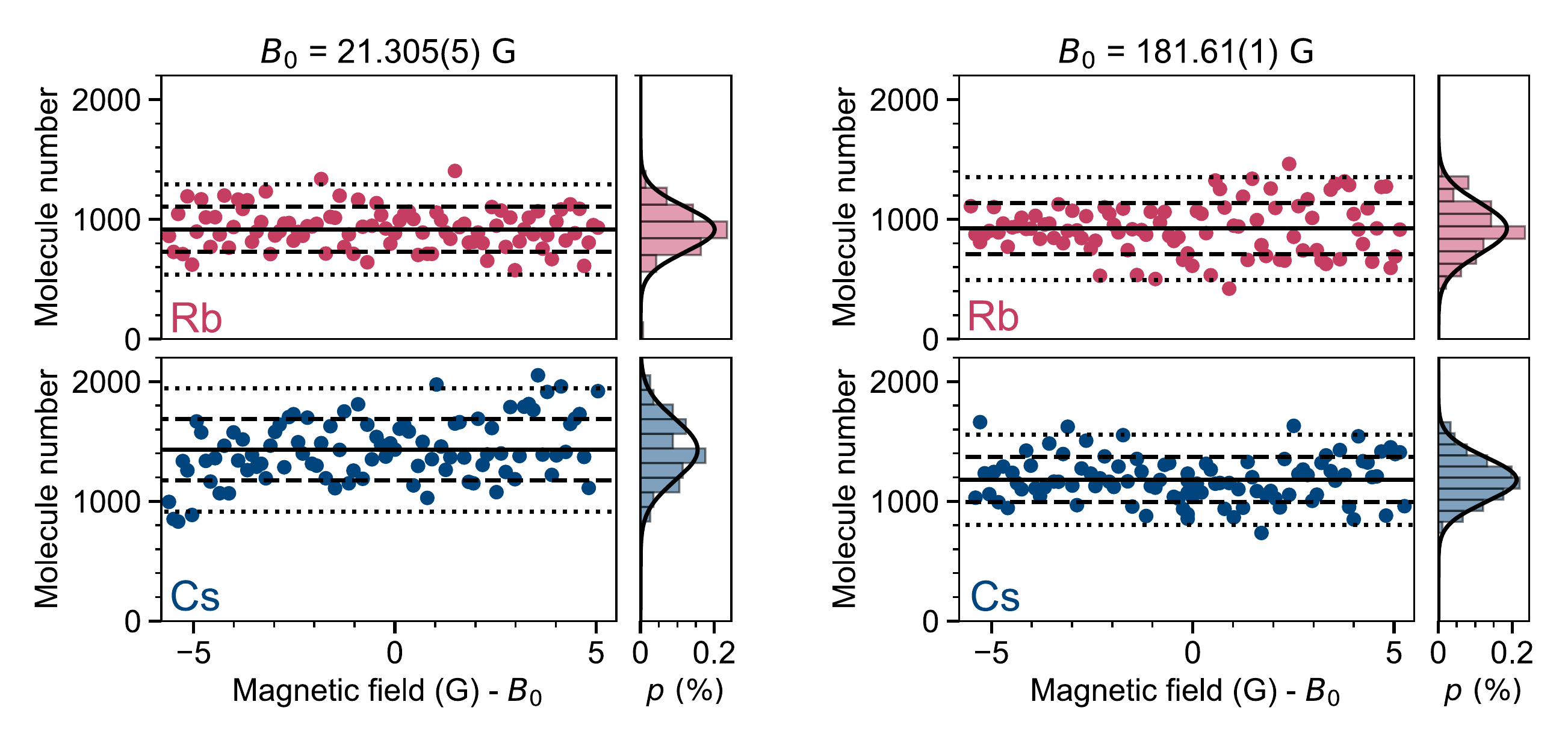}
    \caption{The number of RbCs molecules remaining in $(n=0,m_{f,\mathrm{RbCs}}=5)_0$ after a fixed hold time with Rb atoms in the ($f_\mathrm{Rb}=1, m_{f,\mathrm{Rb}}=1$) state (upper panels) and Cs atoms in the $(f_\mathrm{Cs}=3,m_{f,\mathrm{Cs}}=3)$ state (lower panels) at 21.3\,G (left panels) and 181.6\,G (right panels). In each case the magnetic field is scanned $\pm5$\,G around the centre field $B_0$ in $\sim0.1$~G steps. The atom-molecule mixture is held in the trap for 10\,ms (6\,ms for the data with Rb at $\sim181.6$\,G). The mean value for each panel is shown as a solid horizontal line, 1$\sigma$ and 2$\sigma$ intervals are shown as dashed and dotted lines respectively. The data are also shown as a histogram and compared to a normal distribution, normalised to give the probability density $p$. We see no significant variation from normally distributed noise. 
    }
    \label{fig:NoResonances}
\end{figure}

Finally, motivated by the observation of Feshbach resonances in $^{23}$Na$^{40}$K + $^{40}$K collisions \cite{Yang2019}, we check for the possibility of resonant behaviour close to the magnetic fields at which we perform our experiments. To do this, we hold the atom-molecule mixture in the trap for a fixed period of time (6 to 10\,ms) and vary the magnetic field by $\pm5$\,G in steps of $\sim0.1$\,G, as shown in Fig.~\ref{fig:NoResonances} for molecules prepared in $(0,5)_0$. To determine if there is any variation from background we make a histogram of the results and extract a mean and standard deviation. For all four measurements we observe that only $\sim5\%$ of the points are outside the interval defined by the mean $\pm2\sigma$, as would be expected for normally distributed noise. Additionally, we see no obvious large dips in the molecule number that would indicate the presence of resonances.

\subsection{RbCs + Cs in an intensity-modulated trap}

It is possible that the dominant loss process for atom + molecule collisions is the same fast optical excitation of two-body collision complexes that we have observed for molecule + molecule collisions~\cite{Gregory2019, Gregory2020}. In the RRKM limit, the laser-free lifetime of the RbCs + Cs collision complex in the dark is expected to be a factor of $2\times 10^4$ shorter \cite{Christianen2019DOS,Frye2021} than the lifetime of $\sim0.5$\,ms for the (RbCs)$_2$ complex. If the lifetime of the atom-molecule complex is so short, we may expect that there is not enough time for significant laser excitation before the complex dissociates, or that at least this process may not be saturated. 
However, recent experiments~\cite{Nichols2021} studying nonreactive collisions of KRb with Rb found a photon-free lifetime for the complex of 0.39(6)\,ms, which is 5 orders of magnitude longer than the RRKM prediction, along with evidence for loss associated with optical excitation of the complexes. The reason for such long-lived complexes in KRb+Rb and whether optical excitation of complexes can contribute significantly to loss in collisions between other combinations of atoms and molecules are open questions.

\begin{figure}[t]
    \centering
        \includegraphics[width=0.5\textwidth]{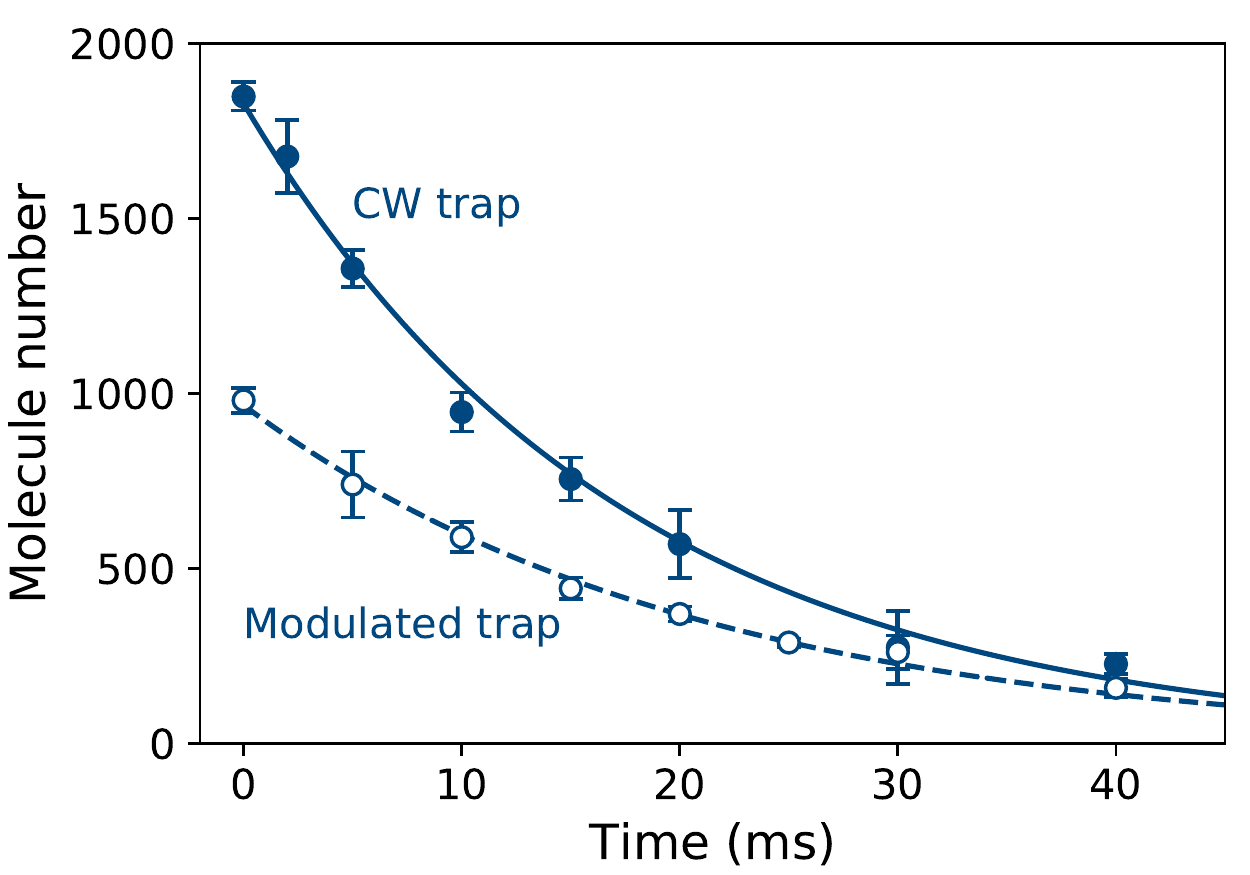}
    \caption{Molecule loss in a mixture of RbCs in the $(n=0,m_{f,\mathrm{RbCs}}=5)_0$ state and Cs in the $(f_\mathrm{Cs}=3,m_{f,\mathrm{Cs}}=3)$ state at 181.6\,G confined in CW (filled circles) and intensity-modulated (empty circles) traps. The lines show fitted exponential decay curves with $1/e$ lifetimes of 17.4(1.2)\,ms in the CW trap and 21.0(1.5)\,ms in the modulated trap.}
    \label{fig:CWvsTimeAvg}
\end{figure}

To investigate this possibility, we have measured the lifetime of RbCs molecules, in the ground state $(0,5)_0$, in the presence of Cs atoms, with and without 1\,kHz square-wave intensity modulation, in the \mbox{$\lambda=1064$\,nm} optical trap described in section~\ref{sec:MolMol}. As for the experiments on RbCs alone, the square-wave modulation is set such that the atom-molecule mixture spends 75\% of each cycle in the dark. The peak intensity in the modulated trap is 4 times that of the CW trap, so that the trap frequencies and trap depths are the same for both trap configurations. 

A comparison of the molecule lifetimes for the CW and modulated traps is shown in Fig.~\ref{fig:CWvsTimeAvg}. We find two-body rate coefficients of $k_\mathrm{mod}=2.1(6)\times10^{-11}$\,cm$^{3}$\,s$^{-1}$ for the modulated trap, and $k_\mathrm{CW}=2.6(6)\times10^{-11}$\,cm$^{3}$\,s$^{-1}$ for the CW trap. As such, there is no statistically significant difference in the loss rates. 

\begin{figure}[t]
    \centering
        \includegraphics[width=0.7\textwidth]{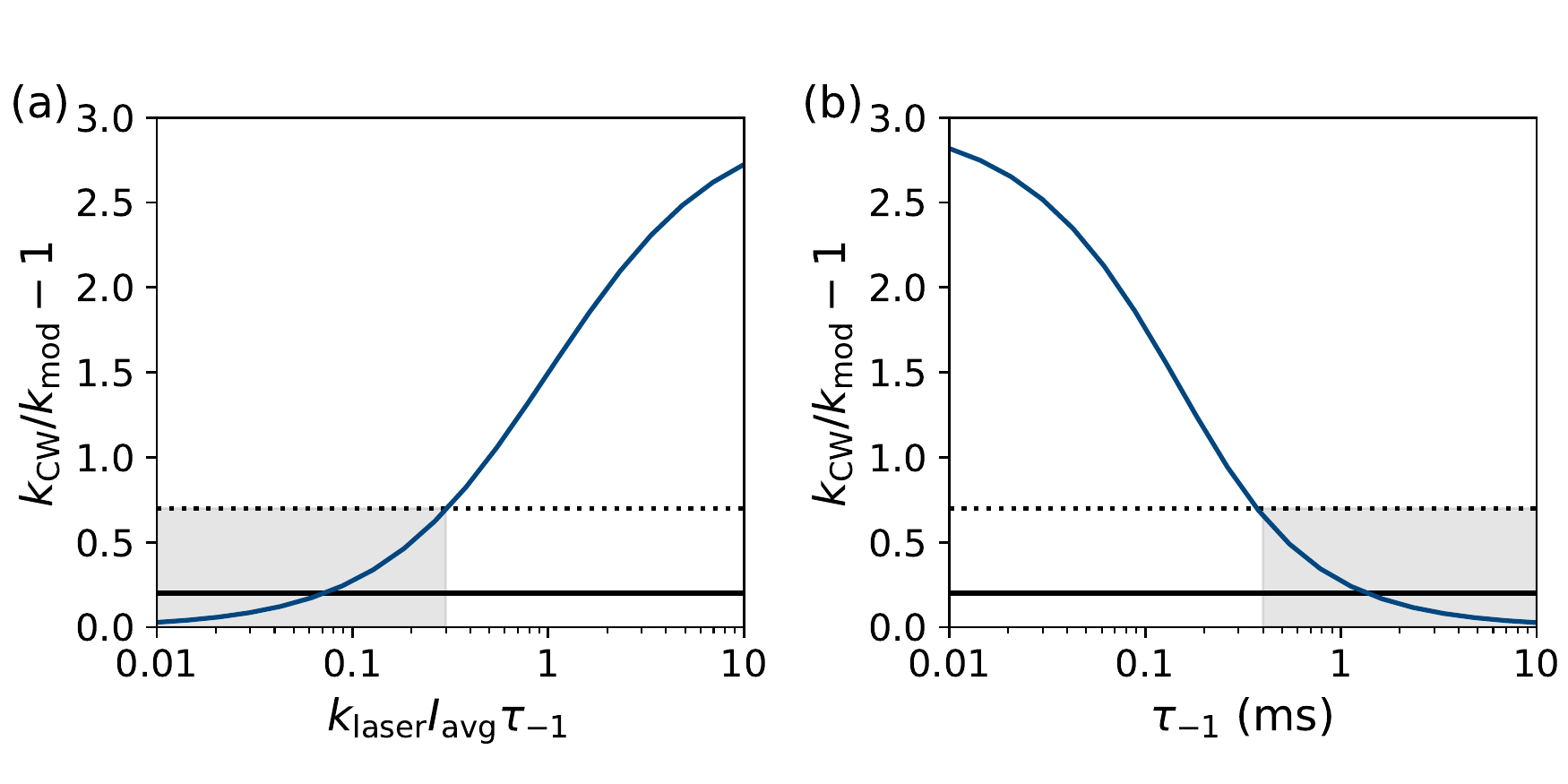}
    \caption{Suppression of the loss of molecules due to atom-molecule collisions in an intensity-modulated trap. We model the variation in the molecular density as described by Eq.~\ref{eq:AtMolRateEquations}, and extract the ratio of loss rates for CW and modulated traps  $(k_\mathrm{CW}/k_\mathrm{mod}-1)$. As presented, our results are independent of the values of $k^\mathrm{a,m}_2$ and $n_\mathrm{a}$ used in the model. The modulation frequency of the trap is fixed at 1\,kHz. We assume no inelastic decay such that the lifetime of the complex $\tau_\mathrm{c}=\tau_{-1}$. (a)~Suppression of loss for unsaturated optical excitation of complexes as a function of $k_\mathrm{laser} I_\mathrm{avg} \tau_{-1}$, in the limit that $\tau_{-1}\ll t_\mathrm{dark}$. (b)~Suppression of loss for saturated optical excitation of complexes as a function of $\tau_{-1}$, in the limit that $k_\mathrm{I}I_\mathrm{avg} \gg 1/\tau_{-1}$. For both calculations, the experimentally measured ratio $k_\mathrm{CW}/k_\mathrm{mod}-1=0.2(5)$ is shown by the horizontal solid line, with the $1\sigma$ uncertainty indicated by the dotted line. The shaded regions show the parameter space which is consistent with our experiments at the 68\% confidence level. }
    \label{fig:AtMolModel}
\end{figure}

To examine our results in the context of optical excitation of atom+molecule complexes, we construct the rate-equation model 
\begin{equation}
    \dot{n}_\mathrm{m} = -k^\mathrm{a,m}_{2}n_\mathrm{a}n_\mathrm{m} + \frac{1}{\tau_\mathrm{-1}}n_\mathrm{c}, \\
    \dot{n}_{c} = +k^\mathrm{a,m}_{2}n_\mathrm{a}n_\mathrm{m} - \frac{1}{\tau_\mathrm{-1}} n_\mathrm{c} - k_\mathrm{laser} I(t) n_\mathrm{c}.
\label{eq:AtMolRateEquations}
\end{equation}
This describes the loss of molecule density due to the atom-molecule collisions and the dissociation and optical excitation of the associated atom-molecule collision complexes. We neglect the formation (and loss) of complexes resulting from molecule-molecule collisions owing to the much longer associated timescale. Additionally, there are no inelastic decay channels for the state combination used in the measurement, such that the lifetime of the complex $\tau_\mathrm{c}=\tau_{-1}$. The predictions of this model are shown in Fig.~\ref{fig:AtMolModel} and lead to two interpretations of the experimental observations.

In Fig.~\ref{fig:AtMolModel}(a) we show the predicted ratio $(k_\mathrm{CW}/k_\mathrm{mod}-1)$
as a function of $k_\mathrm{laser} I_\mathrm{avg} \tau_{-1}$ under the assumption that the lifetime of the complex is much shorter than the trap dark time ($\tau_{-1}\ll t_\mathrm{dark}$) so that the suppression of loss in the modulated trap is independent of the trap modulation frequency. Here $I_\mathrm{avg}$ is the average trap intensity experienced by the molecules in either trap. For modulation where the trap light is off for 75\% of each cycle, we expect $k_\mathrm{CW}/4<k_\mathrm{mod}<k_\mathrm{CW}$ such that the ratio $(k_\mathrm{CW}/k_\mathrm{mod}-1)$ can take values between 0 and 3. The ratio extracted from our experiments is $(k_\mathrm{CW}/k_\mathrm{mod}-1)=0.2(5)$. This is indicated by the horizontal solid and dashed lines in Fig.~\ref{fig:AtMolModel}. The grey shaded region in Fig.~\ref{fig:AtMolModel}(a) indicates the range $k_\mathrm{laser} I_\mathrm{avg} \tau_{-1}<0.3$ consistent with our experimental results at the 68\% confidence level. This corresponds to our results being consistent with optical excitation that is not saturated. This could be due to the predicted short lifetime of the complex, but also depends on the unknown laser scattering rate for the atom-molecule complexes.

In Fig.~\ref{fig:AtMolModel}(b) we examine an alternative interpretation of the experimental results, namely that the lack of suppression in the loss results from the formation of complexes with long photon-free lifetimes. In this case, we assume the optical excitation of the complexes is saturated $k_\mathrm{laser}I_\mathrm{avg}\gg1/\tau_{-1}$. Little or no suppression can then occur if the dark time is not sufficiently long for a significant number of complexes to decay before the next bright trap pulse. Fig.~\ref{fig:AtMolModel}(b) shows the prediction of our model in this limit as a function of $\tau_{-1}$. Our results are consistent with a lifetime of the complex $\tau_{-1}>0.4$\,ms, again at the 68\% confidence level. This would be similar to the recently measured lifetime for complexes in KRb+Rb collisions~\cite{Nichols2021}, but is 5 orders of magnitude longer than the RRKM prediction.

\section{Conclusion}
We have studied collisional loss of optically trapped RbCs molecules alone and in mixtures with Rb and Cs atoms. For RbCs molecules alone, we have demonstrated that collisional loss may be partially suppressed by modulating the intensity of the optical trap, such that the molecules spend 75\% of each modulation cycle in the dark.
For molecules in the spin-stretched absolute ground state, the results confirm that optical excitation of long-lived two-molecule complexes plays a dominant role in the collisional loss. However, the suppression is diminished for molecules in higher-energy hyperfine states. This may indicate changes in the effective density of states or competition from other collisional loss mechanisms such as spin exchange. 

For RbCs molecules prepared in either RbCs+Rb or RbCs+Cs mixtures, we have demonstrated that the collisional loss shows second-order kinetics and we have extracted two-body loss rate coefficients. We have interpreted these using a model based on quantum defect theory. The resulting loss parameters are well below the universal limit, and are similar for reactive collisions with Rb and nonreactive collisions with Cs. For the nonreactive collisions, we have compared the loss rate for mixtures in CW and modulated traps. We observe no significant change in the loss rate associated with the trap light being switched off. 

Understanding ultracold molecular collisions with diatomic molecules remains an important frontier for quantum state-controlled chemistry, with many fundamental questions currently unanswered. The inherent complexity of these systems presents many challenges ahead.

\section*{Acknowledgements}
This work was supported by U.K. Engineering and Physical Sciences Research Council (EPSRC) Grants EP/P01058X/1 and EP/P008275/1.

\section*{References}
\providecommand{\newblock}{}

\end{document}